\documentclass{elsart}
\usepackage{amssymb,epsfig,graphicx}
\begin{document}
\begin{frontmatter}

\title{Structure, Transport and Magnetic properties  in La$_{2x}$Sr$_{2-2x}$Co$_{2x}$Ru$_{2-2x}$O$_{6}$}

\author[gu]{P. S. R. Murthy}, \author[gu]{K. R. Priolkar\corauthref{krp}}\ead{krp@unigoa.ac.in},
\author[tifr]{P. A. Bhobe}, \author[barc]{A. Das}, \author[gu]{P. R. Sarode} and \author[tifr]{A. K. Nigam}
\corauth[krp]{Corresponding author}
\address[gu]{Department of Physics, Goa University, Goa, 403 206 India.}
\address[tifr]{Tata Institute of Fundamental Research, Homi Bhabha Road, Mumbai, 400 005 India}
\address[barc]{Solid State Physics Division, Bhabha Atomic Research Centre, Trombay, Mumbai 400 085 India}

\begin{abstract}
The perovskite solid solutions of the type La$_{2x}$Sr$_{2-2x}$Co$_{2x}$Ru$_{2-2x}$O$_{6}$ with 0.25 $\leq$ x $
\leq $ 0.75 have been investigated for their structural, magnetic and transport properties. All the compounds
crystallize in double perovskite structure. The magnetization measurements indicate a complex magnetic ground
state with strong competition between ferromagnetic and antiferromagnetic interactions. Resistivity of the
compounds is in confirmation with hopping conduction behaviour though differences are noted especially for $x$ =
0.4 and 0.6. Most importantly, low field (50Oe) magnetization measurements display negative magnetization during
the zero field cooled cycle. X-ray photoelectron spectroscopy measurements indicate presence of
Co$^{2+}$/Co$^{3+}$ and Ru$^{4+}$/Ru$^{5+}$ redox couples in all compositions except $x$ = 0.5.  Presence of
magnetic ions like Ru$^{4+}$ and Co$^{3+}$ gives rise to additional ferromagnetic (Ru-rich) and antiferromagnetic
sublattices and also explains the observed negative magnetization.
\end{abstract}

\begin{keyword}

\PACS{72.15.Jf; 81.30.Kf; 75.50.Cc}
\end{keyword}
\end{frontmatter}

\section{Introduction}
LaSrCoRuO$_6$ is a double perovskite whose magnetic properties critically depend on  cationic order, charge
balance and complex magnetic interactions between two transition metal ions \cite{bos,mam,ramu}. Its crystal
structure is composed of corner-shared CoO$_6$ and RuO$_6$ octahedra arranged in a pseudocubic array in the
rocksalt arrangement. It is a semiconductor with ideal valence states HS Co$^{2+}$ (3d$^7$ high-spin
configuration) and Ru$^{5+}$ (4d$^3$) \cite{kim}. Magnetically these compounds are reported to be
antiferromagnetic with two magnetic face centered cubic (fcc) sublattices consisting of Co and Ru. Both the
sublattices order with type II antiferromagnetic structure which would mean that the spins in [111] planes in
succession Co-Ru-Co-Ru alternate as +/+/-/ -. This marginalizes the Co-O-Ru nearest neighbor interactions and the
ordering is governed by a competition between linear Co-O-Ru-O-Co and 90$^\circ$ Co-O-O-Co antiferromagnetic
exchange paths \cite{bos,dlouha}. The degree of ordering is known to affect the magnetic and transport properties
due to changes in magnetic interactions and in cationic valence. Effects of anti-site disorder on the magnetic
and transport properties due to La or Sr doping in LaSrCoRuO$_6$ have been investigated \cite{bos,dlouha,tomes}.
The change of composition (La or Sr doping) introduces mobile electrons in La richer samples or holes in Sr
richer samples.

Another possible way of modifying magnetic and transport properties is by forming the solid solutions of
antiferromagnetic LaCoO$_3$ and ferromagnetic SrRuO$_3$. These perovskite solid solutions of the form
Sr$_{1-x}$La$_x$Ru$_{1-x}$Co$_x$O$_3$ will have a strong interplay of cationic order, charge balance and complex
magnetic interactions between the two B-site cations. In SrRuO$_3$ the $4d$ electrons of the low spin Ru$^{4+}$
ions occupy the narrow $\pi^*$ band near Fermi level \cite{gb1}. The lower $3d$ energy levels of Co$^{3+}$ causes
a charge transfer from $4d$ Ru$^{4+}$ to $3d$ Co$^{3+}$ \cite{mam}. However, Co can have various electronic
states including high spin (HS) Co$^{2+}$, Co$^{3+}$ and Co$^{4+}$, intermediate spin (IS) Co$^{3+}$ and
Co$^{4+}$ and low spin (LS) Co$^{3+}$ and Co$^{4+}$ \cite{gb,potze,saitoh,caci,rad}. This complicates the
situation giving rise properties like localized magnetic moment of Co \cite{14}, negative magnetoresistance
\cite{15}. In case of  Sr$_{1-x}$La$_x$Ru$_{1-x}$Co$_x$O$_3$, a complete charge transfer occurs at $x$ = 0.5. At
this composition the CoO$_6$ and RuO$_6$ octahedra align themselves in a pseudocubic array in the rocksalt
arrangement forming the arch-type "double perovskite" structure.

The charge transfer between Ru and Co in LaSrCoRuO$_6$ is very sensitive to local atomic structure such as cation
order \cite{bos,ramu,tomes}. Any disturbance in this cation order leads to compensation of antiferromagnetic
interactions by the ferromagnetic interactions most likely associated with Ru-O-Ru interactions. The LaCoO$_{3}$
substituted SrRuO$_{3}$, has been known to exhibit large local magnetic moment arising due to Co spin polarizing
the itinerant electrons of SrRuO$_3$ \cite{mam}. However, the delicate charge balance achieved in solid solutions
by conversion of  Co$^{3+}$ to Co$^{2+}$ and oxidation Ru$^{4+}$  to Ru$^{5+}$ due to formation of double
perovskite structure has not been addressed. More importantly the magnetic interactions at play as the system
transforms from a ferromagnetic ground state to an antiferromagnetic one is far from being clearly understood. It
is with this aim the present investigation is proposed. This paper describes detailed investigations carried out
on the structural, magnetic and transport properties of solid solutions of SrRuO$_3$ and LaCoO$_3$ which form
double perovskite compounds of the type La$_{2x}$Sr$_{2-2x}$Co$_{2x}$Ru$_{2-2x}$O$_{6}$, where 0.25 $ \leq $ x $
\leq $ 0.75.

\section{Experimental}
Polycrystalline samples of La$_{2x}$Sr$_{2-2x}$Co$_{2x}$Ru$_{2-2x}$O$_{6}$, $0.25  \leq  x  \leq 0.75$ were
synthesized by solid state reaction method by taking pre-dried stoichiometric amounts of La$_{2}$O$_{3}$,
SrCO$_{3}$, Co(NO$_{3}$)$_{2}$.6H$_{2}$O and RuO$_{2}$. These starting powders were ground thoroughly, pressed
into pellets and heated for a total of 48 hrs, at 1300$^\circ$C with three intermediate regrinding steps. All
samples were deemed to be phase pure, as X-ray diffraction (XRD) data collected on a Rigaku X-ray diffractometer
in the range of $ 18^\circ \le 2\theta \le 80^\circ$ using CuK$\alpha$ radiation showed no impurity reflections.
The diffraction patterns were Rietveld refined using FULLPROF suite and structural parameters were obtained. DC
magnetization (M) was measured, both as a function of temperature and magnetic field using the Quantum Design
SQUID magnetometer (MPMS-5S). Magnetization as a function of temperature, M(T) was measured in an applied field
of 50 Oe and 1000 Oe in the temperature range of 5K to 300K. The sample was initially cooled from 300K to 5K in
zero applied field and the data was recorded while warming up to 300K in the applied magnetic field (referred to
as ZFC curve) and subsequent cooling (referred to as FC curve) back to 5K. Magnetization as a function of field
was measured under sweep magnetic fields up to $\pm$5T at various temperatures. Before each M(H) was recorded,
the sample was warmed to 300K and cooled back to the desired temperature. Neutron diffraction (ND) measurements
were performed at room temperature (RT) and 20K and a wavelength of 1.24\AA~ using powder diffractometer at
Dhruva, Trombay. X-ray photoelectron spectra at Co (2p) and Ru (3d) levels were recorded using Thermo Fisher
Scientific Multilab 2000 (England) instrument with Al K$_\alpha$ radiation (1486.6 eV). The binding energies
reported here is with reference to graphite at 284.5 eV having an accuracy of $\pm$0.1 eV.

\section{Results}

\begin{figure}[c]
\centering
\includegraphics[scale=0.5]{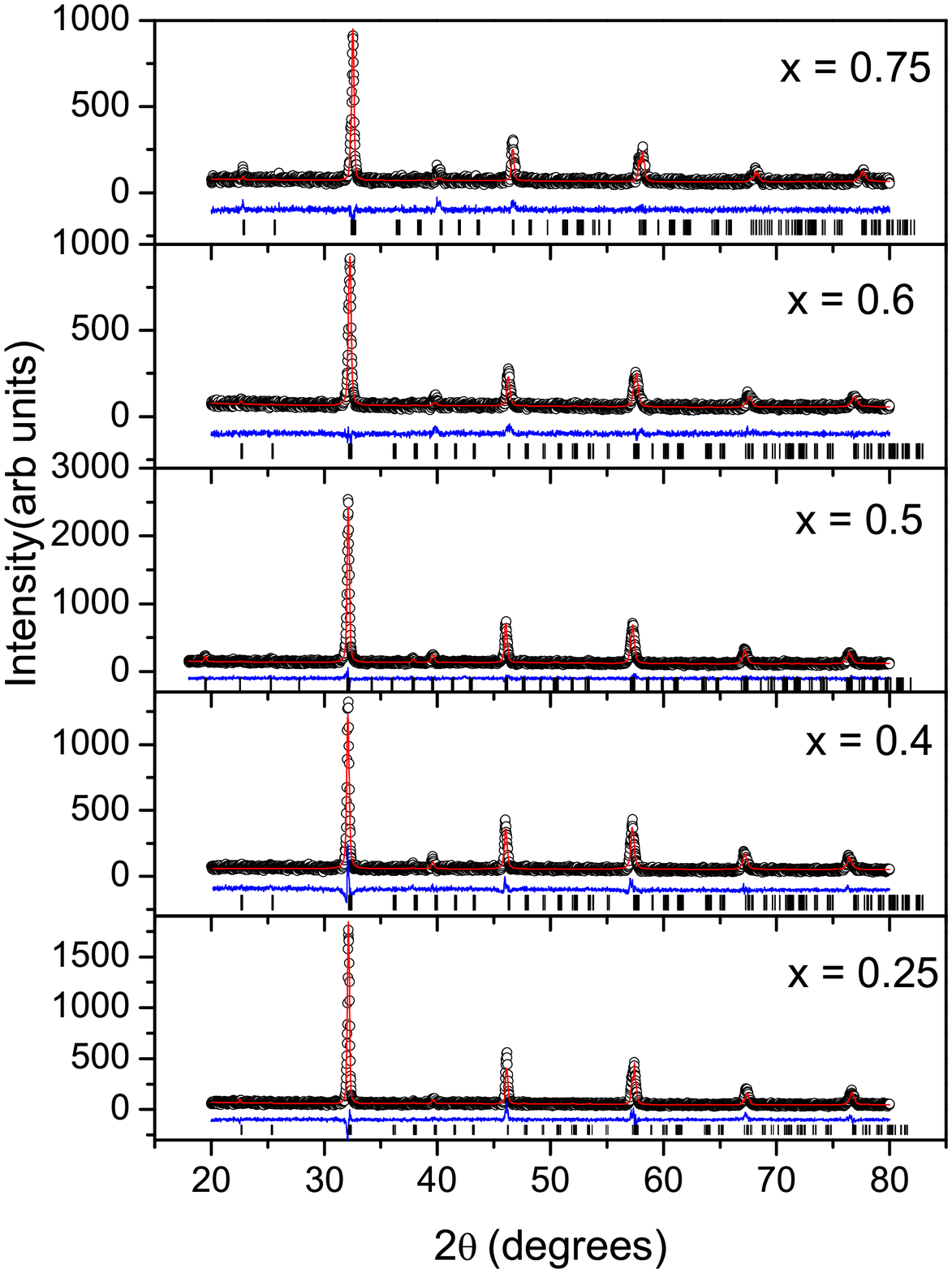}
\caption{\label{SeriesIIXRD} Rietveld refined XRD patterns of La$_{2x}$Sr$_{2-2x}$Co$_{2x}$Ru$_{2-2x}$O$_{6}$.
The open circles show the observed counts and the continuous line passing through these counts is the calculated
profile. The difference between the observed and calculated patterns is shown as a continuous line at the bottom
of the two profiles. The calculated positions of the reflections are shown as vertical bars.}
\end{figure}

The Rietveld refined XRD patterns for all the compounds studied here are presented in Fig. \ref{SeriesIIXRD}. The
samples crystallize in the P2$_{1/n}$ monoclinic structure with a initial increase followed by a decrease beyond
$x$ = 0.5 of cell volume as LaCoO$_{3}$ is added to SrRuO$_{3}$ to form a solid solutions. It may be mentioned
here that the compounds with values of $x < 0.25$ and $x > 0.75$ have Pbnm and R$\bar3$c structures respectively
and hence were not studied as they cannot be classified as double perovskites. Rietveld refinement of the XRD
patterns were carried out with the P2$_{1/n}$ space group wherein the La/Sr occupy the 4e site with the
fractional coordinates (0.00125,0.00774, 0.2463), Co is at 2c (0.5, 0, 0.5), Ru is at 2d (0.5, 0, 0) and the
oxygen atoms occupy three sites, viz,(0.2491, 0.2566, 0.0295); (0.2207, -0.2233, 0.0295) and (-0.06418, 0.4995,
0.2507). The scale factor, back ground parameters, cell parameters, Co and Ru site occupancies along with
instrumental broadening , totalling 17 parameters were refined in that order to obtain a good fit. As can be seen
from Table \ref{as} the B-site disorder was found to be least for $x$ = 0.5 which also happens to be the
stoichiometric double perovskite LaSrCoRuO$_6$. Interestingly the disorder or the deviation from expected
occupancy is highest for $x$ = 0.4 and 0.6. This aspect needs more attention and perhaps magnetic and transport
properties will shed light on this. The crystallographic parameters obtained from the above refinements along
with the Curie-Weiss parameters calculated from magnetization measurements  are all summarized in Table
\ref{riet}.

\begin{table}
\caption{\label{as} Expected (E) and refined (R) occupancies of Co and Ru for different values of
La$_{2x}$Sr$_{2-2x}$Co$_{2x}$Ru$_{2-2x}$O$_{6}$.} \centering
\begin{tabular}{cc|cc|cc|cc|cc|cc}
\hline \multicolumn{2}{c}{$x \to $} & \multicolumn{2}{c}{0.25} & \multicolumn{2}{c}{0.4} & \multicolumn{2}{c}{0.5} & \multicolumn{2}{c}{0.6} & \multicolumn{2}{c}{0.75} \\
\hline
sites &  atoms & E & R & E & R & E & R & E & R & E & R\\
(${1\over 2},0,{1\over 2}$) & Co & 0.25 & 0.19 & 0.4& 0.26 & 0.5 & 0.49 & 0.5 & 0.36 & 0.5 & 0.45\\
& Ru & 0.25 & 0.31 & 0.1 & 0.24 & -- & 0.01 & -- & 0.14 & -- & 0.05\\
(${1\over 2}),0,0$) & Ru & 0.5 & 0.44 & 0.5 & 0.36 & 0.5 & 0.49 & 0.4 & 0.26 & 0.25 & 0.20 \\
& Co & -- & 0.06 & -- & 0.14 & -- & 0.01 & 0.1 & 0.24 & 0.25 & 0.30\\
\hline
\end{tabular}
\end{table}

\begin{table*}
\caption{\label{riet} Unit cell parameters obtained from Rietveld refinement and Curie-Weiss parameters
calculated from magnetization measurements at 1000 Oe for La$_{2x}$Sr$_{2-2x}$Co$_{2x}$Ru$_{2-2x}$O$_{6}$.
Numbers in parentheses are uncertainty in the last digit.} \centering
\begin{tabular}{cccccccc}
\hline $x$ & $a$(\AA) & $b$(\AA) & $c$(\AA) & $\beta^\circ$ & V (\AA$^3$) & $\Theta_{CW}(K)$ &
$\mu_{eff}(\mu_B)$ \\
0.25 & 5.5577(4) & 5.5715(6) & 7.8409(9) & 90.05(2) & 242.80(4) & 62.8(2) & 3.49(2)\\
0.40 & 5.5750(3) & 5.5733(3) & 7.8683(9) & 90.25(2) & 244.47(4) & 49.5(2) & 3.44(2)\\
0.50 & 5.5847(3) & 5.5591(3) & 7.8674(9) & 90.05(2) & 244.25(4) & -49.0(2) & 3.87(2)\\
0.60 & 5.5626(3) & 5.5287(3) & 7.8245(9) & 90.08(2) & 240.63(4) & -20.8(2) & 3.86(1)\\
0.75 & 5.4824(3) & 5.5310(5) & 7.7697(5) & 89.93(3) & 235.60(3) & -24.8(4) & 3.82(1)\\
\hline
\end{tabular}
\end{table*}

\begin{figure}[c]
\centering
\includegraphics[scale=0.5]{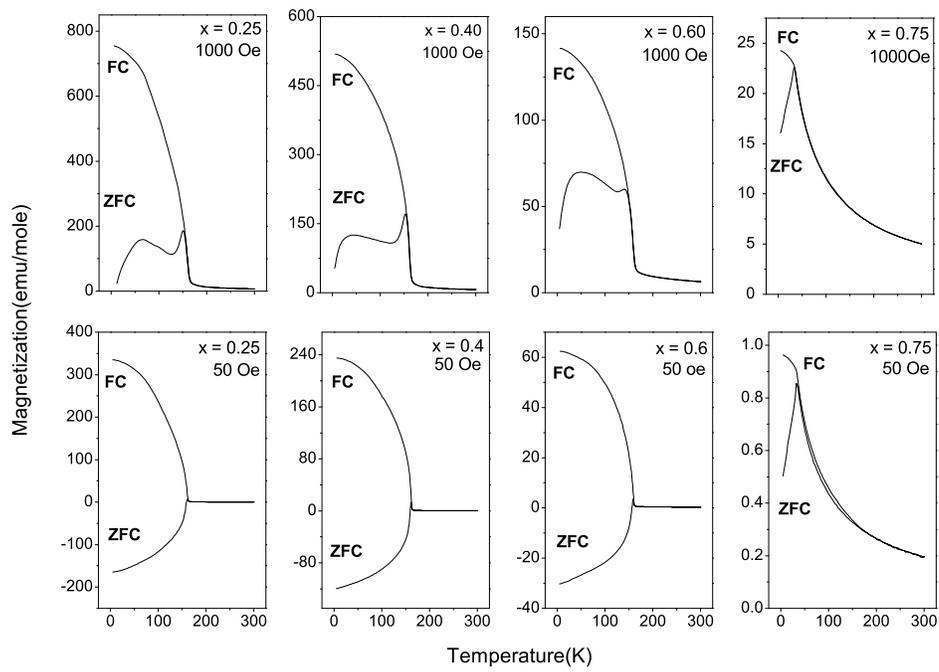}
\caption{\label{SeriesIIMT} Magnetization as a function of temperature recorded in applied fields of 1000 Oe
(upper panel) and 50 Oe (lower panel) in La$_{2x}$Sr$_{2-2x}$Co$_{2x}$Ru$_{2-2x}$O$_{6}$.}
\end{figure}

Magnetization measurements performed in applied fields of 1000 Oe and 50 Oe during the ZFC and FC cycles for $x$
= 0.25, 0.4, 0.6 and 0.75 samples are presented in Fig. \ref{SeriesIIMT}. In the case of $x$ = 0.25, 0.4 and 0.6
there is a wide difference in magnetization recorded during the ZFC and FC cycles. The ZFC magnetization for
these three samples, with increasing temperature increases sharply culminating into a broad hump centered around
50K. It decreases slightly with further rise in temperature before increasing sharply resulting in a peak at
about 150K. The FC magnetization, on the other hand decreases continuously to about 167K and then settles down
into a low value giving an impression of a ferro to para transition. The wide difference in the magnetization
between the ZFC and FC cycle indicates a complex magnetic ground state. It may also be noted that the
magnetization (emu/mole) value at 5K decreases with increasing La and Co content.  Further with increasing $x$
the irreversibility between the ZFC and FC curves is also seen to decrease until at $x$ = 0.75. For this
composition there is a very little difference between ZFC and FC magnetization curves and the sample seems to
order antiferromagnetically at T$_{N}$ = 34 K.

\begin{figure}[c]
\centering
\includegraphics[scale=0.5]{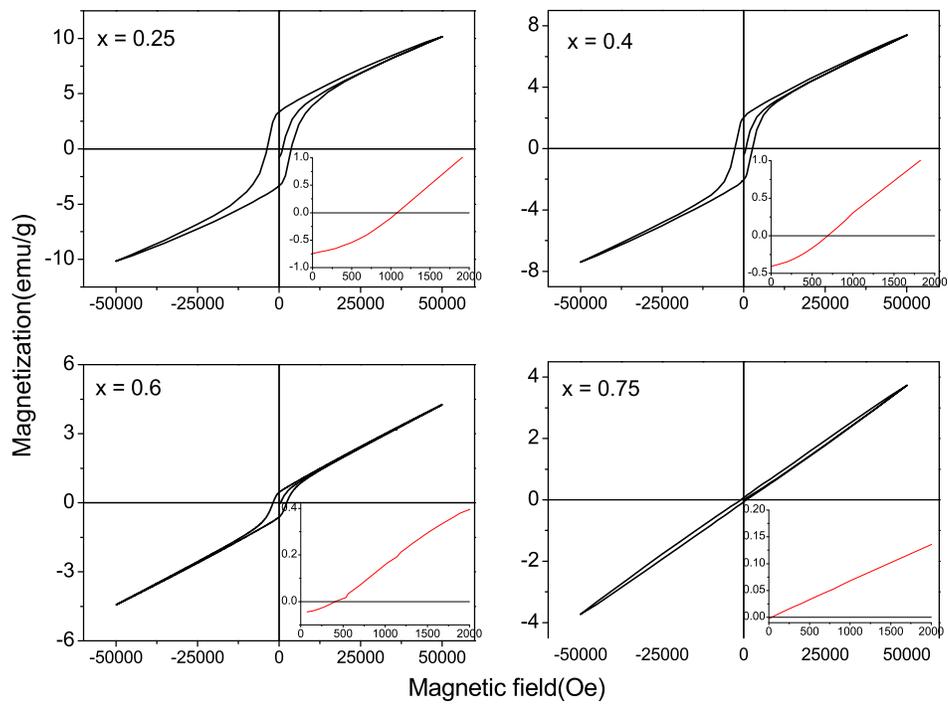}
\caption{\label{SeriesIIMH}Isothermal magnetization curves recorded in the field interval of $\pm$5T at 5K for
La$_{2x}$Sr$_{2-2x}$Co$_{2x}$Ru$_{2-2x}$O$_{6}$.}
\end{figure}

The low field magnetization measurements are also in agreement with the above. The interesting point however is
the observation of negative values of magnetization for three compositions viz, $x$ = 0.25, 0.4 and 0.75 during
ZFC cycle. Such negative values of magnetization were also seen in case of thermally disordered LaSrCoRuO$_6$
\cite{ramu}. It may be noted here that all precautions were taken to ensure that negative values of magnetization
are not due to remanent field of superconducting magnet of the SQUID magnetometer and this procedure has been
described earlier \cite{ramu}. Furthermore, as has been described later, the initial magnetization curves
recorded for these samples at 5K also exhibit negative magnetization for lower values of field (see insets of
Fig. \ref{SeriesIIMH}). In case of $x$ = 0.75 however, the magnetization was all along positive even during the
ZFC cycle. The negative magnetization could be ascribed to presence of two magnetic sub-lattices which order in
such a way as to cancel the magnetization of each other and in low fields align the net magnetic moment in a
direction opposite to the applied field. The two magnetic sublattices could be conjunctured to be ferromagnetic
Ru$^{4+}$-O-Ru$^{4+}$ and antiferromagnetic Co$^{2+}$-O-Ru$^{5+}$ and/or Co$^{3+}$-O-Co$^{3+}$.

\begin{figure}[c]
\centering
\includegraphics[scale=0.75]{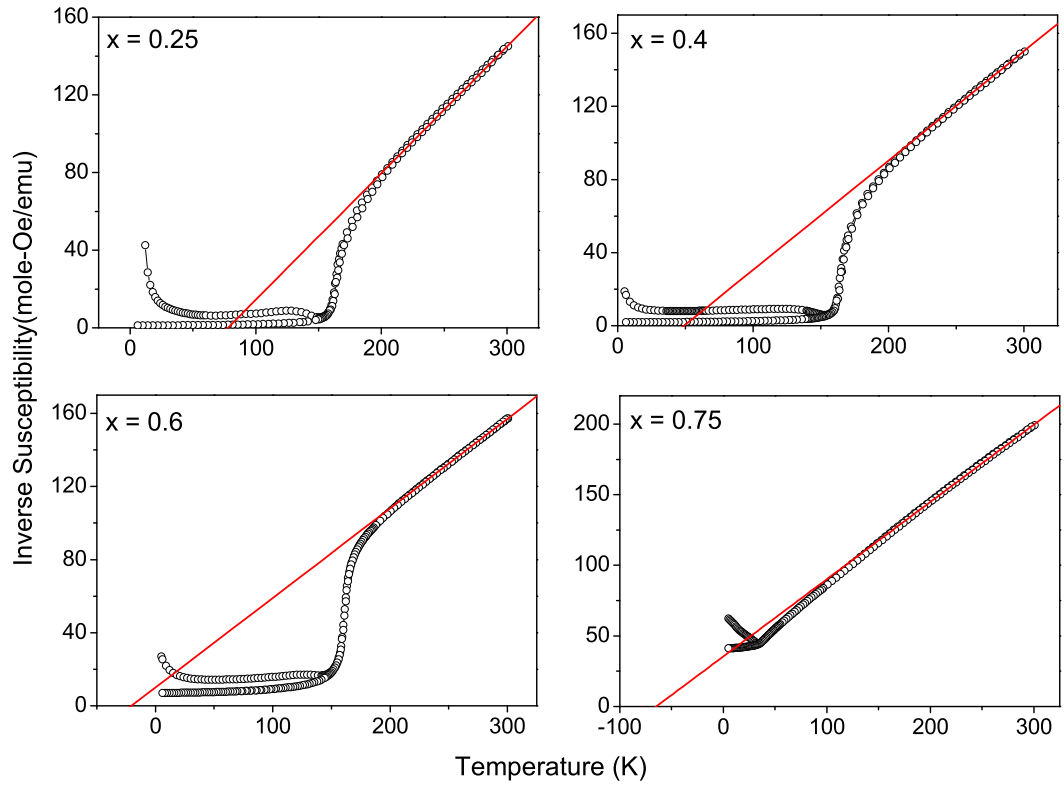}
\caption{\label{Suscp}Plot of inverse magnetic susceptibility (H/M) as a function of temperature calculated using
magnetization data recorded in applied field of 1000 Oe in La$_{2x}$Sr$_{2-2x}$Co$_{2x}$Ru$_{2-2x}$O$_{6}$.}
\end{figure}

This fact will be more clear from the values of effective paramagnetic moment $\mu_{\rm{eff}}$ and Curie-Weiss
temperature $\Theta_{CW}$ obtained from the high temperature magnetization behaviour. Curie-Weiss analysis has
been employed to examine the behaviour of high temperature magnetization. The plots of the inverse susceptibility
(1/$\chi$ = H/M) versus temperature are presented in Fig \ref{Suscp}. For x = 0.75, the inverse susceptibility
appears to vary linearly with temperature in the range 40K $<$ T $<$ 300K but a Curie-Weiss fit to the data
indicates that there is a deviation from the linear fit below 160K. This is a characteristic progressive
suppression of the spin-spin interactions as temperature decreases due to spin-orbit coupling \cite{jin}. The
antiferromagnetic order below 34K can then be attributed to ordering of Co spins via non-magnetic RuO$_{6/2}$
bridges \cite{kim1}. Therefore the data in the range 180K $<$ T $<$ 300K was fitted to the Curie-Weiss law and
the values of $\mu_{eff}$ and $\Theta_{CW}$ were obtained. Likewise for all the other compounds too Curie-Weiss
fitting was performed in the high temperature range (180K to 300K) and $\mu_{eff}$ and $\Theta_{CW}$ were
calculated. These parameters are listed in Table \ref{riet}. It can be seen that while the $\mu_{eff}$ varies
only in a small range between 3.44 to 3.87 $\mu_B$, $\Theta_{CW}$ shows a parabolic variation with $x$ and
changes its sign from negative for Co rich compounds to positive for Ru rich compositions. Negative $\Theta_{CW}$
indicates presence of stronger antiferromagnetic interactions while positive $\Theta_{CW}$ means stronger
ferromagnetic interactions. In case of $x$ = 0.5, the $\mu_{eff}$ value agrees very well with the calculated spin
only moment value of Co$^{2+}$ and Ru$^{5+}$ indicating formation of a well ordered double perovskite. Another
point to be noted is very large negative value of $\Theta_{CW}$. This is an indicator of strong antiferromagnetic
interactions and indeed this compound is reported to order antiferromagnetically at $T_N$ = 80K \cite{bos}.  In
other cases, the Ru rich compositions obviously have higher amount of Ru$^{4+}$/Ru$^{5+}$ leading to stronger
ferromagnetic interactions and positive $\Theta_{CW}$ while the Co rich compounds have stronger antiferromagnetic
interactions arising from higher amounts of Co$^{2+}$/Co$^{3+}$ ions. However, for all these compositions,
$\mu_{eff}$ values reported in Table \ref{riet} can only be obtained by considering the presence of
Ru$^{4+}$/Ru$^{5+}$ and Co$^{2+}$/Co$^{3+}$ redox couples. Even in case of $x$ = 0.75 small deviation between ZFC
and FC curves is seen below 160K (see Fig. \ref{SeriesIIMT}). The presence of these magnetic ions results in
formation of more than one magnetic sublattices giving rise to complex magnetic behaviour.

\begin{figure}[c]
\centering
\includegraphics[scale=0.5]{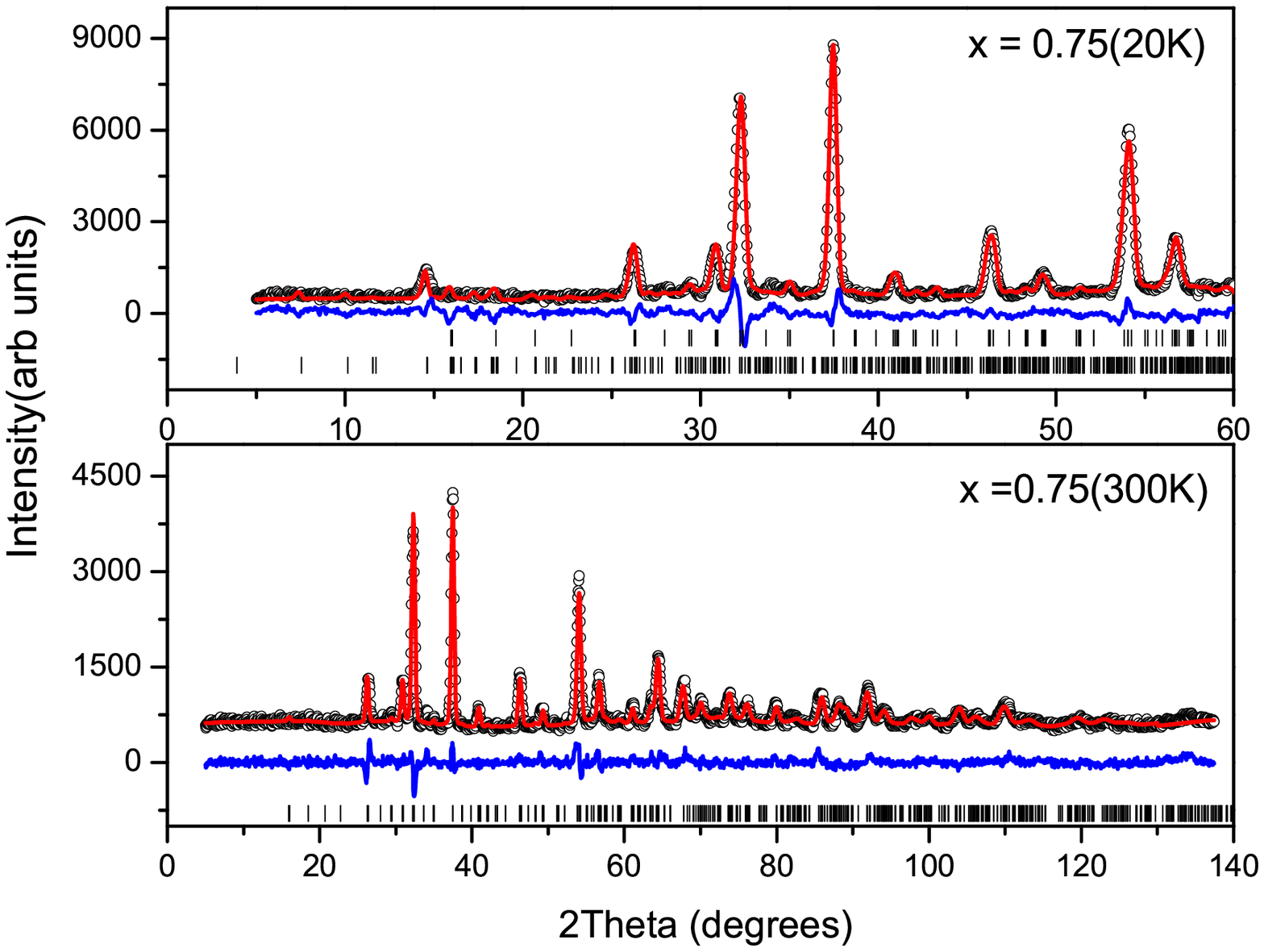}
\includegraphics[scale=0.5]{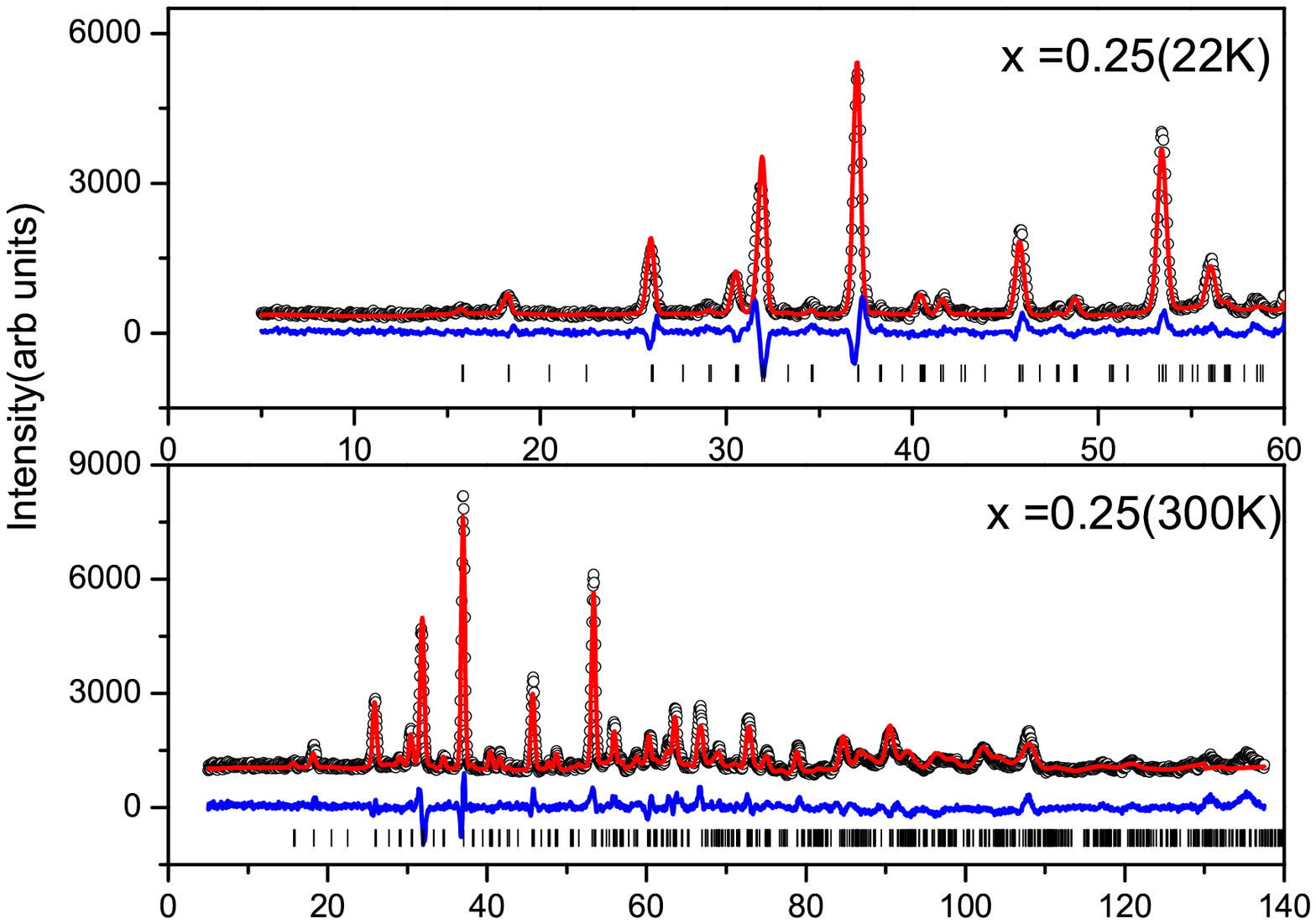}
\caption{\label{nd} Observed (circles), calculated (line) ND patterns recorded at 300K and 20K in case of
La$_{2x}$Sr$_{2-2x}$Co$_{2x}$Ru$_{2-2x}$O$_{6}$ for x = 0.25 and x = 0.75. The data at 20K is shown in limited
range for clarity. The continuous line at the bottom is the difference line between observed and calculated
data.}
\end{figure}

To establish the nature of magnetic order, ND patterns were recorded at low temperature (20K) and 300K for the
two end members $x$ = 0.75 and 0.25.  The Rietveld refined ND patterns at 300K for both these compounds are
presented in Fig. \ref{nd}. The parameters obtained from Rietveld refinement agree well with those obtained from
XRD studies. The low temperature (20K) data shown in limited range in Fig. \ref{nd} indicates extra superlattice
reflections due to antiferromagnetic ordering in $x$ = 0.75. These reflections can be accounted for by an
antiferromagnetic alignment of Co and Ru spins with a propagation vector along the $k = {1\over 2}, 0, {1\over
2}$ with respect to crystallographic axis. The magnetic arrangement is the same as determined for $x$ = 0.5 in Ref. \cite {bos}. As per this model, the antiferromagnetic alignment is of type II which means that antiparallel spins are related by (${1\over 2},{1\over 2},{1\over 2}$) translation operation. The refined spin moments for Ru and Co which were taken to be equal, were $\mu_x$ = 0.67(3)$\mu_B$, $\mu_z$ = 0.36(6)$\mu_B$ and resultant $\mu$ = 0.69(2)$\mu_B$. No long range magnetic order is visible in case of Ru rich composition ($x$ = 0.25) indicating that the sharp rise in magnetization at about 160K is due to short range ferromagnetic correlations. The short range ferromagnetic correlations could be due to Ru-O-Ru linkages intervened by Co ions. These ferromagnetic correlations are present along with antiferromagnetic interactions as evidenced by the wide separation between ZFC and FC curves for this sample.

A further confirmation of the presence of competing magnetic interactions is obtained from isothermal magnetic
response recorded for all the samples at various temperatures in the field range of $\pm$50KOe. Fig.
\ref{SeriesIIMH} presents the isothermal magnetization curves for four samples ($x$ = 0.25, 0.4, 0.6 and 0,75) at
5K. It can be seen from this figure that for $x$ = 0.75, the magnetization exhibits a strong field dependency and
almost no hysteresis which is typical of an antiferromagnet. This is in agreement with antiferromagnetic ordering
seen in M v/s T and ND measurements. On the other hand for all other compositions the isothermal magnetization
exhibits a clear ferromagnetic hysteresis loop that rides on an antiferromagnetic (linear) background. Therefore
the observed magnetic behavior can be ascribed to the presence of competing ferromagnetic and antiferromagnetic
interactions. It is also be noticed that with increase in Ru content, the area under the hysteresis loop
increases implying the strengthening of ferromagnetic interactions. The initial magnetization curves shown as
insets in Fig. \ref{SeriesIIMH} make it clear that the negative magnetization seen in the low field ZFC curves is
indeed an intrinsic property of the materials studied here. Further it can also be seen that with the increase in
$x$, magnetization turns positive at lower and lower values of applied field.

\begin{figure}[c]
\centering
\includegraphics[scale=0.5]{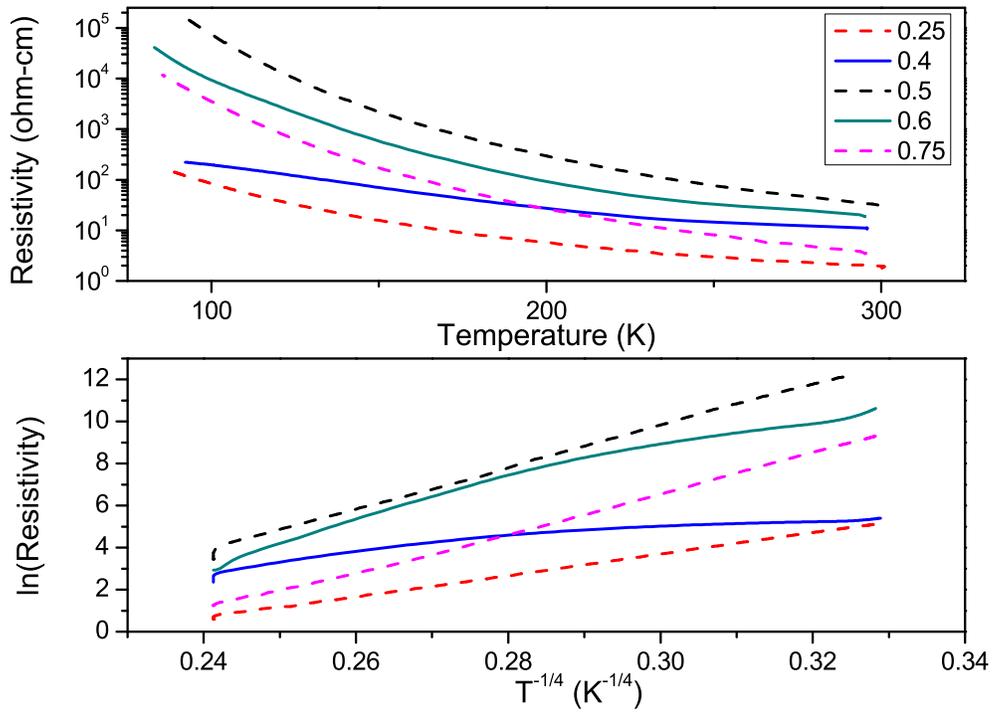}
\caption{\label{SeriesIIRest}Resistivity of La$_{2x}$Sr$_{2-2x}$Co$_{2x}$Ru$_{2-2x}$O$_{6}$ as a function of
temperature (upper panel). The lower panel shows a plot of $\log\rho$ versus T$^{-1/4}$ for all compositions.}
\end{figure}

Presence of Co$^{2+/3+}$ and Ru$^{4+/5+}$ redox couples that give rise to complex magnetic behaviour will also
affect the transport properties of the compounds. It may be mentioned here that while SrRuO$_3$ has metallic
conductivity \cite{srruo3}, LaCoO$_3$ exhibits semiconducting behaviour at temperatures below 300K \cite{lacoo3}.
A plot of $\rho$ versus temperature for La$_{2x}$Sr$_{2-2x}$Co$_{2x}$Ru$_{2-2x}$O$_{6}$ with $0.25 \le x \le
0.75$ is presented in Fig.\ref{SeriesIIRest}. All the samples show semiconducting behaviour. It may be noted that
the ordered composition ($x$ = 0.5) has the highest magnitude of resistivity due to absence of any Ru-O-Ru type
conducting paths. Since the resistivity of this ordered sample follows Mott's variable range hopping (VRH)
behaviour, $\log\rho$ versus T$^{-1/4}$ has been plotted in the lower panel of fig. \ref{SeriesIIRest} for all
compositions. It can be seen that along with $x$ = 0.5, $x$ = 0.25 and 0.75 samples show a linear behaviour
indicating conduction to be in accordance with Mott's VRH law. Resistivity behaviour for $x$ = 0.4 and 0.6
compounds however, is quite different. Here the $\rho$ varies in a very narrow range and exhibits a hump at about
160K which coincides with the sharp rise in magnetization data of these samples. Rietveld analysis have also
shown maximum B-site occupancy disorder for the same two compositions and therefore could be linked to the
presence of more number of Ru-O-Ru conducting paths.

\begin{figure}[c]
\centering
\includegraphics[scale=0.25]{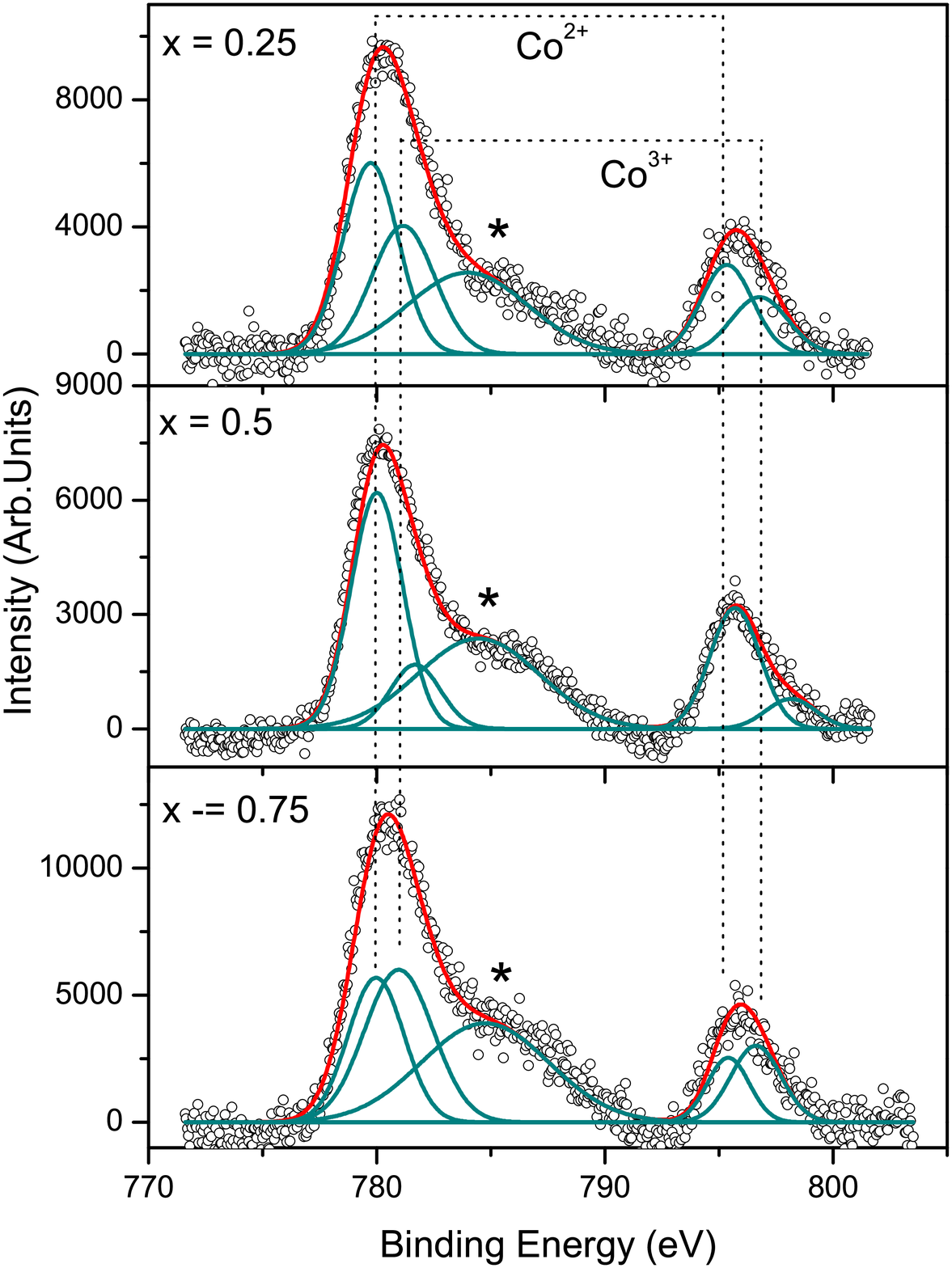}
\includegraphics[scale=0.25]{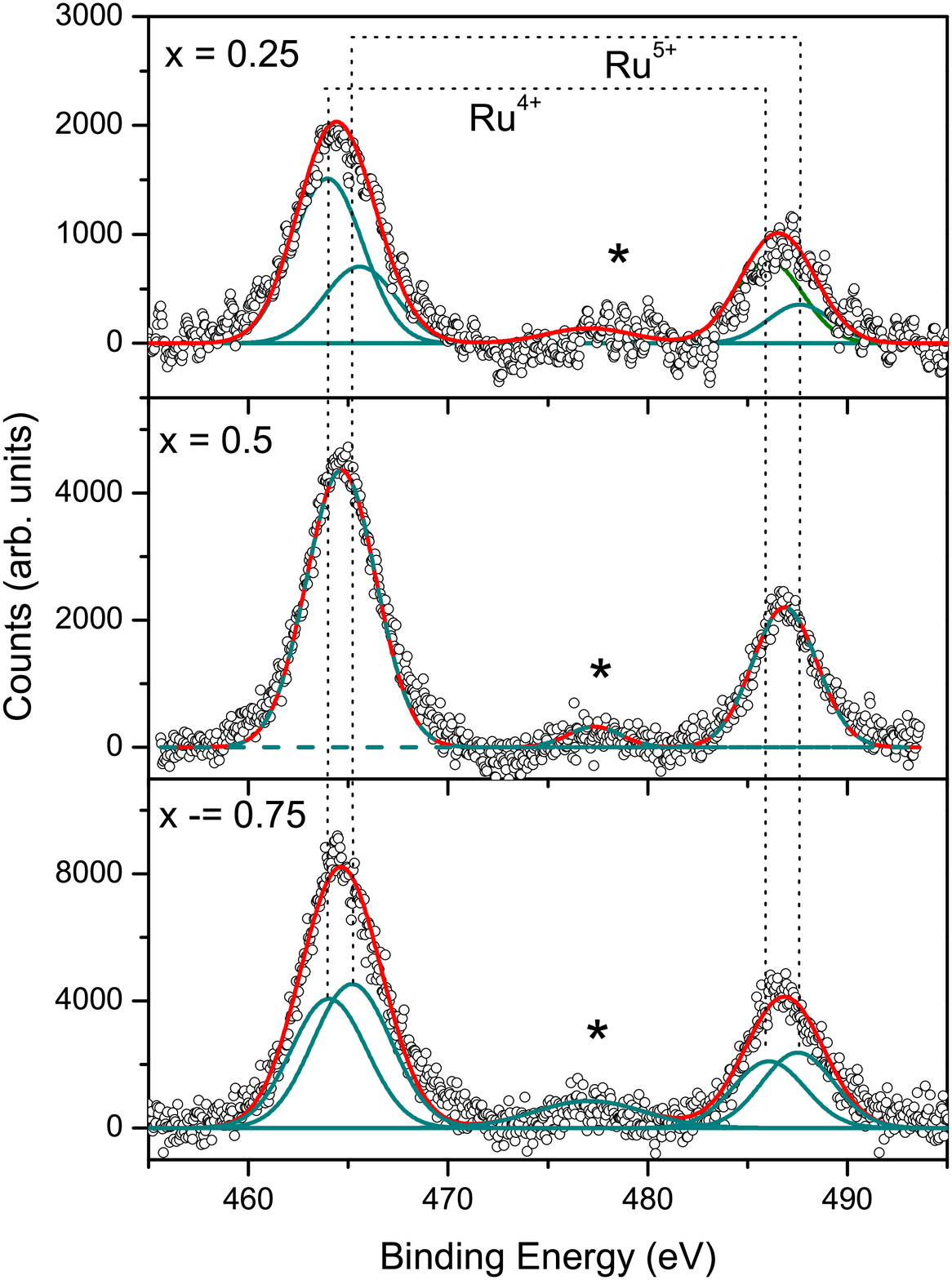}
\caption{\label{XPS} Co(2p) and Ru(3p) core level spectra along with fitted curves for
La$_{2x}$Sr$_{2-2x}$Co$_{2x}$Ru$_{2-2x}$O$_{6}$ compounds.}
\end{figure}

All the above properties hint at presence of Co$^{2+}$/Co$^{3+}$ and Ru$^{4+}$/Ru$^{5+}$ pairs in
La$_{2x}$Sr$_{2-2x}$Co$_{2x}$Ru$_{2-2x}$O$_6$ compounds in different proportion. Core level XPS of Co and Ru can
give an indication of valency of these ions and therefore a measure of such proportion. Fig \ref{XPS} presents
the background subtracted Co and Ru core level spectra for $x$ = 0.25, 0.5 and 0.75. Both Co $2p$ and Ru $3p$
spectra show two clear peaks due to spin orbit splitting and the associated satellite peaks (marked as *). In the
case of Co, the main peaks are separated by a spin orbit splitting of about 15eV while in the case of Ru this
splitting is about 22eV \cite{xps}. In the case of $x$ = 0.25 and 0.75, the main peaks are broadened as compared
to $x$ = 0.5 and show considerable spectral weight on the higher energy side especially in case of Co $2p$.
Therefore the spectra have been fitted with two Gaussians perhaps corresponding to Co$^{2+}$ and Co$^{3+}$
species. Likewise the Ru $3p$ spectra also shows a presence of two types of Ru ions which are most likely to be
Ru$^{4+}$ and Ru$^{5+}$ species. In case of $x$ = 0.5, Co $2p$ and Ru $3p$ spectra can be well represented by a
single gaussian which can be attributed to divalent Co and pentavalent Ru respectively. Further in case of $x$ =
0.25 which has majority Ru content, the intensity of the peak corresponding to Ru$^{4+}$ is higher than that of
the peak corresponding to Ru$^{5+}$ while the $x$ = 0.75 sample shows higher amounts of Co$^{3+}$ species as
compared to Co$^{2+}$. This is as expected, the unexpected however is the corresponding contents of
Co$^{2+}$/Co$^{3+}$ in $x$ = 0.25 and Ru$^{4+}$/Ru$^{5+}$ in $x$ = 0.75. Using the percentage concentration
ratios of Co$^{2+}$/Co$^{3+}$ and Ru$^{4+}$/Ru$^{5+}$ obtained from area under the respective Gaussian's an
attempt was made to calculate the spin only magnetic moments for the two compositions. The calculations were made
assuming S = 3/2 for Co$^{2+}$ and Ru$^{5+}$, S = 1 for Co$^{3+}$ and S = 2 for Ru$^{4+}$. The calculated values
were found to be $3.5 \mu_B$ in case of $x$ = 0.75 and $3.25 \mu_B$ in case of $x$ = 0.25. Although slightly
lower, they seem to agree with the trend reported in Table \ref{riet}.

\section{Discussion}
In the case of ordered double perovskite, LaSrCoRuO$_6$, the presence of highly acidic Ru$^{5+}$ stabilizes the
high spin Co$^{2+}$ and thereby suppressing various electronic transitions that Co ion can have. However, when
the order is disturbed, Co-O-Co and Ru-O-Ru linkages are formed and the trivalent state of Co and tetravalent
state of Ru are favoured. Such has been the case in La$_{2-x}$Sr$_{x}$CoRuO$_6$ \cite{tomes}.

In La$_{2x}$Sr$_{2-2x}$Co$_{2x}$Ru$_{2-2x}$O$_{6}$, present studies indicate that with the addition of LaCoO$_3$
to SrRuO$_3$ leads to stabilization of double perovskite phase is seen in a broad concentration region of $0.25
\le x \le 0.75$. These double perovskites, apart from Co$^{2+}$-O-Ru$^{5+}$ interactions will also contain
Co-O-Co and Ru-O-Ru linkages. This is amply clear from the magnetic and transport properties described above. As
mentioned above, these linkages favour trivalent and tetravalent states for Co and Ru respectively. Further,
under such conditions it is well known that HS and IS states of Co$^{3+}$ are promoted \cite{knizek}. Presence of
HS/IS Co$^{3+}$ ions will give rise to additional magnetic interactions. Further Ru$^{4+}$-O-Ru$^{4+}$ is known
to be ferromagnetic and metallic in conduction \cite{djsingh}. The observed competition between ferromagnetic and
antiferromagnetic interactions can be therefore attributed to presence of Ru$^{4+}$ and Co$^{3+}$ ions. The
presence of these ions along with occupancy disorder leads to formation of Ru rich and Co rich magnetic
sublattices which order ferromagnetically and antiferromagnetically respectively. This is well supported by the
decrease in magnetization values with increasing Co content and sharper rise of magnetization in case of Ru rich
compositions.  Only when there is a complete ordering of the B-site cations, Co$^{2+}$ and Ru$^{5+}$ ions are
stabilized as nearest neighbours  and long range antiferromagnetic order is established. The antiferromagnetic
order seen in $x$ = 0.75 can be attributed to antiferromagnetic cobalt ordering via non-magnetic RuO$_6/2$
bridges \cite{kim1}.

The presence of more than one magnetic sublattices also explains the negative magnetization observed in ZFC cycle
of low field magnetization data. The ferromagnetic sublattice formed due to presence of tetravalent Ru ions
orders at around 160K. This polarizes the paramagnetic Co spins in a direction opposite to the applied field
leading to magnetic compensation and negative values of magnetization. With a decrease in Ru content, the
ferromagnetic sublattice becomes weaker and the field required to reverse the magnetization to positive values
also decreases. This can be very clearly seen from the initial magnetization curves presented in Fig.
\ref{SeriesIIMH}.

\section{Conclusion}
\sloppy In summary, the structural, transport and magnetic properties in
La$_{2x}$Sr$_{2-2x}$Co$_{2x}$Ru$_{2-2x}$O$_{6}$ have been studied. Double perovskite structure with space group
P2$_{1/n}$ is stabilized over a wide composition range from $x$ = 0.25 to 0.75. With the increase in Co content,
ferromagnetic interactions are found to weaken and at $x$ = 0.75 the compound orders antiferromagnetically at
$T_N$ = 34K. This interplay of ferromagnetic and antiferromagnetic interactions is attributed to presence of
Ru$^{4+}$/Ru$^{5+}$ and Co$^{2+}$/Co$^{3+}$ redox couple in all the compounds. The only exception to this is the
ordered compound $x$ = 0.5 wherein Co and Ru exist in divalent and pentavalent states respectively representing
an archtype double perovskite. The presence of different magnetic sublattices leads to magnetic compensation and
negative magnetization. This can be explained by polarization of paramagnetic Co spins by the  ferromagnetic
Ru$^{4+}$ sublattice in a direction opposite to applied field.

\section*{Acknowledgements}
KRP and PRS would like to thank Department of Science and Technology (DST), Government of India for financial
support under the project No. SR/S2/CMP-42. KRP and PSRM acknowledges support from UGC-DAE Consortium for
Scientific Research, Mumbai Centre for financial support under CRS-M-126. Authors also thank Prof. M. S. Hegde
for useful discussions.


\newpage

\end{document}